\documentclass[9pt,a4 paper]{article}
\usepackage{amssymb}
\usepackage{amsmath}
\usepackage[dvips]{graphicx}
\begin{document}
\thispagestyle{empty}

\noindent {\textbf{ Holographic dark energy models with statefinder diagnostic in modified $f(R,T)$ gravity  }}

\vspace{1cm}

\noindent  \textbf{C. P. Singh \footnote{Corresponding author}, \;\textbf{Pankaj Kumar\;$^{2}$ }}\\

\vspace{0.5cm}

\noindent{ $^{1,2}$Department of Applied Mathematics,\\
 Delhi Technological University (Formerly Delhi College of
 Engineering)\\
 Bawana Road, Delhi-110 042, India.}\\
\texttt{ $^1$cpsphd@rediffmail.com  \\
$^2$pankaj.11dtu@gmail.com}\\

\vspace{1.5cm}

\noindent {\textbf{Abstract}}  We study non-viscous and viscous holographic dark energy models for a homogeneous and isotropic flat Friedmann-Robertson-Walker Universe in $f(R,T)$ gravity. We find that the Hubble horizon as an IR cut-off is suitable for both the models to explain the recent accelerated expansion of the Universe. The cosmological parameters like deceleration parameter and statefinder parameters are discussed in each model. In non-viscous model a constant deceleration parameter is found which shows that there is no phase transition. The constraints on the parameters are obtained to analyze the fixed point values of statefinder parameters of SCDM and $\Lambda$CDM models. We know that the phase transition is required to explain the accelerated expansion of the Universe and this is possible if both the parameters would be time-dependent. Therefore, we extend our analysis to viscous holographic dark energy model to investigate whether this viscous model with the same IR cut-off could be helpful to find the phase transition. We find that this model gives a time-dependent deceleration parameter which achieves a smooth phase transition of the Universe. We also find the time-varying  statefinder pair which matches with $\Lambda$CDM model. We plot the trajectories in $r-s$ and $r-q$ plans to discriminate our model with the existing dark energy models and obtain the quintessence like behavior for the suitable values of parameters.

\pagebreak

\pagestyle{myheadings}
\noindent\textbf{1 Introduction}\\

\noindent It is strongly believed that the Universe has entered a phase of the accelerated expansion which has been confirmed by the recent observations like supernovae Ia [1, 2], cosmic microwave background radiation [3, 4], baryon acoustic oscillation [5] and Planck data [6]. Within the framework of general relativity (GR), the cause of the acceleration can  be attributed to the existence of a mysterious component of the Universe dubbed as ``dark energy" (DE), which makes up $\sim$ $70\%$ of the total cosmic energy in the Universe. The $\Lambda$CDM model presents the simplest and most successful description of the recent accelerated expansion scenario and accommodates the observations very well. Despite of many attractive features, it has some theoretical problems like fine-tuning and cosmic coincidence problems [7-9]. To overcome from these problems, a number of dynamical dark energy models such as scalar field (quintessence, phantom, k-essence, etc.) models [10-13], chaplygin gas models [14], holographic dark energy (HDE) models [15-17], etc. have been explored in the literature . \\
\indent In the recent years, the HDE models have been emerged as a viable candidates to explain the problems of modern cosmology. The HDE models explain the recent accelerated expansion as well as the coincidence problem of the Universe [18, 19]. The concept of HDE is based on the holographic principle proposed by 't Hooft [20] and found it's roots in the quantum field theory. Cohan et al. [21] have shown that in the quantum field theory, the formation of black hole set a limit which relates UV cut-off $\Lambda$ to IR cut-off $L$. According to the authors, the quantum zero-point energy $\rho_{h}=\Lambda^4$ of a system of size $L$ should not exceed the mass of a black hole of the same size, i.e., $L^3\rho_{h}\leq L M_{p}^2$, where $M_{p}=(8\pi G)^{-1/2} $ is the reduced Planck mass. In a paper [18], Li has taken the largest allowed $L$ to saturate this inequality and thus obtained dark energy density of the Universe $\rho_{h}=3c^2M_{p}^2L^{-2}$, known as HDE density. In the formalism of HDE, the Hubble horizon is a most natural choice for the IR cut-off, but it leads to a wrong equation of state (EoS) of dark energy [22]. However, Pav\'{o}n and Zimdhal [19], and Banerjee and Pav\'{o}n [23] have shown that the viable EoS of dark energy could be achieved by taking the interaction between HDE and dark matter (DM).\\
\indent On the other hand, the modified theories of gravity such as $f(R)$ gravity [24, 25], $f(G)$ gravity [26, 27], $f(R, G)$ gravity [28, 29], etc. have also been proposed to explain the recent accelerated expansion of the Universe. The $f(R)$ gravity is one of the simplest and successful modified theories of GR, which fits with the observations very well. Recently, Harko et al. [30] have proposed a new modified gravity theory known as $f(R, T)$ gravity, where $R$ as usual stands for the Ricci scalar and $T$ denotes the trace of energy-momentum tensor. This modified theory presents a maximal coupling between geometry and matter. A number of authors [31-36] have discussed the modified $f(R, T)$ gravity in different context to explain the early and late time acceleration of the Universe. In a recent paper [37], the authors have discussed the viscous cosmology in this theory which shows the recent phase transition of the Universe. The HDE models have not been yet discussed in detail in the framework of f(R,T) gravity. In some papers [38, 39], reconstruction of f(R, T) gravity from HDE and anisotropic model of HDE have been discussed.\\
\indent In this paper, we are interested to discuss HDE with Hubble horizon as an IR cut-off in the framework of $f(R, T)$ gravity. It has been shown in GR and Brans-Dicke theory that the Hubble horizon as an IR cut-off is a suitable candidate to explain the recent accelerated expansion if we consider interaction between HDE and DM [19, 23]. As it is known, the $f(R, T)$ gravity has inbuilt interaction between geometry and matter, therefore, it will be interesting to discuss HDE with Hubble horizon as an IR cut-off in this modified theory. The observations from different probes suggest that the DM is a non-interacting content of the Universe. Therefore, it is assumed that only HDE component of the total matter (HDE +DM) interacts with the geometry of the Universe i.e. $T$ of $f(R, T)$ is the trace of the energy-momentum tensor of HDE only. We show that the Hubble horizon as an IR cut-off is suitable to explain the recent accelerated expansion in this modified gravity theory.\\
 \indent The Hubble parameter $H$ and the deceleration parameter $q$ are well known cosmological parameters which explain the evolution of the Universe. However, these two parameters can not discriminate among various DE models. In this context, Sahni et al. [40, 41] have introduced a new geometrical diagnostic pair $\{r, s\}$, known as statefinder parameters, which is constructed from the scale factor and its derivatives up to the third order. The statefinder pair $\{r, s\}$ is geometrical in the nature as it is constructed from the space-time metric directly. Therefore, the statefinder parameters are more universal parameters to study the DE models than any other physical parameters. In a flat $\Lambda$CDM model, the statefinder pair has a fixed point value $\{r, s\}=\{1, 0\}$. One can plot the trajectories in $r-s$ and $r-q$ planes to discriminate various DE models. We discuss the statefinder diagnostic and obtain the fixed point values of statefinder pair $\{r, s\}=\{1, 1\}$ and $\{r, s\}=\{1, 0\}$ as in the case of SCDM and $\Lambda$CDM models, respectively, under suitable constraints.\\
\indent To be more realistic, the prefect fluid Universe is just an approximation of the viscous Universe. The dissipative processes in the relativistic fluid may be modeled as bulk viscosity. The phenomenon of the bulk viscosity arises in the cosmological fluid when the fluid expands (contracts) to fast due to which the system is out of thermal equilibrium. Then, the effective pressure become negative to restore the thermal equilibrium [42]. Therefore, it is natural to consider the bulk viscosity in an accelerating Universe. It has been shown that inflation and recent acceleration can be explained using the viscous behavior of the Universe, and plays an important role in the phase transition of the Universe [43-50]. The concept of viscous DE has been discussed extensively in the literature [51-53]. Feng and Li [54] show that the age problem of the Ricci dark energy can be alleviated using the bulk viscosity.  Motivated by the above works, we extend our analysis to viscous HDE with the same IR cut-off which gives the recent phase transition of the Universe. We obtain the statefinder parameters for viscous HDE which achieve the value of $\Lambda$CDM model and show consistency with the quintessence model for suitable value of parameters.\\
\indent The paper is organized as follows. In the next section we discuss the formalism of $f(R,T)$ gravity theory and present its field equations. In section 3 we discuss the non-viscous HDE model and find the exact power-law solution of the scale factor which avoids the big bang singularity. We also find the cosmological parameters like deceleration parameter and statefinder parameters  and discuss their behaviors. Section 4 describes the viscous HDE model and its solution. Section 5 presents the summary of our findings.\\

\noindent \textbf{2 The formalism of modified $f(R,T)$ gravity theory}\\

\indent The general form of the Einstein-Hilbert action for the modified $f(R, T)$ gravity in the unit $8\pi G=1$ is as follows [30, 36]:
\begin{equation}
S=\frac{1}{2}\int d^{4}x\sqrt{-g}\;[f(R,T)+2\mathcal{L}_{m}],
\end{equation}
\noindent where $g$ stands for the determinant of the metric tensor $g_{\mu\nu}$, $R$ is the Ricci scalar  and $T$ represents the trace of the energy-momentum tensor, i.e., $T= T^{\mu}_{\mu}$,  while $\mathcal{L}_{m}$ denotes the matter Lagrangian density. The speed of light is taken to be unity. As usual the energy- momentum tensor, $T_{\mu\nu}$ of matter is defined as
\begin{equation}
T_{\mu\nu}=-\frac{2}{\sqrt{-g}}\frac{\delta(\sqrt{-g}\mathcal{L}_{m})}{\delta g^{\mu\nu}}.
\end{equation}
In fact, this modified gravity is the generalization of $f(R)$ gravity and is based on the coupling between geometry and matter. The corresponding field equations have been derived in metric formalism for the various forms of $f(R,T)$. \\
 \indent Varying the action (1) with respect to the metric tensor $g_{\mu\nu}$ for a simple form of $f(R,T)=R+f(T)$, i.e., the usual Einstein-Hilbert term plus an $f(T)$ correction [30, 36] which  modifies the general relativity and represents a coupling with geometry of the Universe, we get the following field equations.
\begin{equation}
R_{\mu\nu}-\frac{1}{2}R g_{\mu\nu}=T_{\mu\nu}- (T_{\mu\nu}+\circleddash_{\mu\nu})f'(T)+ \frac{1}{2}f(T) g_{\mu\nu},
\end{equation}
where a prime denotes derivative with respect to the argument.  The tensor $\circleddash_{\mu\nu}$ in (3) is given by
\begin{equation}
\circleddash_{\mu\nu}=-2T_{\mu\nu}+g_{\mu\nu}\mathcal{L}_m-2g^{\alpha\beta}\frac{\partial^2\mathcal{L}_m}{\partial g^{\mu\nu}\partial g^{\alpha\beta}}.
\end{equation}
The matter Lagrangian $\mathcal{L}_{m}$ may be chosen as $\mathcal{L}_{m}=-p$ [30], where $p$ is the thermodynamical pressure of matter content of the Universe. Now, Eq. (4) gives $\circleddash_{\mu\nu}=-2T_{\mu\nu}-p g_{\mu\nu}$. Using this result, Eq. (3) reduce to
\begin{equation}
R_{\mu\nu}-\frac{1}{2}R g_{\mu\nu}=T_{\mu\nu}+ (T_{\mu\nu}+p\;g_{\mu\nu})f'(T)+ \frac{1}{2}f(T) g_{\mu\nu},
\end{equation}
\noindent which are the field equations of the modified $f(R,T)$ gravity theory.\\
\noindent Here, we are interested to study the behavior of HDE in this modified theory for a spatially homogeneous and isotropic flat Friedmann-Robertson-Walker (FRW) spacetime, which is expressed in comoving coordinates by the line element,
\begin{equation}
ds^{2}=dt^{2}-a^{2}(t)(dx^{2}+dy^{2}+dz^{2}),
\end{equation}
\noindent where $a(t)$ stands for the cosmic scale factor. In what follows we study the non-viscous and viscous HDE models with deceleration parameter and statefinder parameters in $f(R,T)$ gravity theory to describe the recent acceleration.\\

\noindent \textbf{3. Non-viscous holographic dark energy cosmology}\\

In this model, let us consider the Universe filled with HDE plus pressureless DM (excluding baryonic matter), i.e.,
 \begin{equation}
  T_{\mu\nu}=T_{\mu\nu}^h+T_{\mu\nu}^m,
\end{equation}
\noindent where $T_{\mu\nu}^h$ and $T_{\mu\nu}^m$ represent the energy-momentum tensors of HDE and DM, respectively. Many authors have described the recent accelerated expansion by assuming the interaction between HDE and DM in the different theories of gravity. In this paper, instead of taking the interaction between HDE and DM to describe the recent acceleration, we consider that the HDE interacts with the geometry of $f(R, T)$ gravity. This is due to the fact that this modified gravity theory has the interaction between matter and geometry. Therefore, we consider $T=g^{\mu\nu}T_{\mu\nu}^h$ as the trace of energy-momentum tensor of HDE. The generalized Einstein equations (5) yield
\begin{equation}
3H^{2}=\rho_{m}+\rho_{h} + (\rho_{m}+\rho_{h}+p_{h})f'(T)+\frac{1}{2}f(T),
\end{equation}
\begin{equation}
2\dot{H}+3H^{2}=-p_{h}+\frac{1}{2}f(T),
\end{equation}
 where $\rho_{m}$, $\rho_{h}$ and $p_{h}$ denote the energy density of DM, the energy density of HDE and the pressure of HDE, respectively. An overdot denotes the derivative with respect to cosmic time $t$. As the field equations (8) and (9) are highly non-linear, therefore, we assume $f(T)=\alpha\;T$ [see, ref. 30], where $\alpha$ is a coupling parameter.  Now, the field equations (8) and (9) reduce as
\begin{equation}
3H^{2}=\rho_{m}+\rho_{h} + \alpha(\rho_{m}+\rho_{h}+p_{h})+\frac{1}{2}\;\alpha\; T,
\end{equation}
\begin{equation}
2\dot{H}+3H^{2}=-p_{h}+\frac{1}{2}\;\alpha\; T.
\end{equation}
The equation of state (EoS) and the trace of energy-momentum tensor of HDE are given by $p_{h}=w_{h}\rho_{h}$ and $T=\rho_{h}-3p_{h}$, respectively. Now, from (10) and (11), a combined evolution equation for $H$ can be written as
\begin{equation}
2\dot H + (1+\alpha)[(1+w_{h})\rho_{m}+\rho_{h}]=0.
\end{equation}
\indent In the literature, various forms of HDE ( the general form is $\rho_h=3c^2M_{p}^{2}L^{-2}$, where $c^2$ is a dimensionless constant, $M_{p}$ stands for the reduced Planck mass and $L$ denotes the IR cut-off radius) have been discussed depending on the choices of IR cut-off such as Hubble horizon, future event horizon, apparent horizon, Granda-Oliveros cut-off, etc. In this work, we consider the Hubble horizon ($L= H^{-1}$) as an IR cut-off to describe the recent acceleration. The corresponding energy density $\rho_h$  is given by
\begin{equation}
\rho_{h}=3\;c^2H^2.
\end{equation}
Form (10) and (13), the energy density of DM can be written as
\begin{equation}
\rho_{m}=\frac{3(\alpha c^2 w_{h}-3\alpha c^2-2 c^2+2)}{2(1+\alpha)}\;H^2.
\end{equation}
 Using (13) and (14) into (12), we finally get
\begin{equation}
\dot H+\frac{3}{4} (3\alpha c^2 w_{h}+2 c^2 w_{h}-\alpha c^2+2)\;H^2=0,
\end{equation}
which, on solving gives
\begin{equation}
H=\frac{1}{c_{0}+\frac{3}{4} (3\alpha c^2 w_{h}+2 c^2 w_{h}-\alpha c^2+2)t},
\end{equation}
where $c_{0}$ is a positive constant of integration. Eq. (16) can be rewritten as
\begin{equation}
H=\frac{H_{0}}{1+\frac{3 H_{0}}{4} (3\alpha c^2 w_{h}+2 c^2 w_{h}-\alpha c^2+2)(t-t_{0})},
\end{equation}
where $H_{0}$ is the present value of the Hubble parameter at the cosmic time $t=t_{0}$, the time where the HDE starts to dominate. Using the relation $H=\frac{\dot {a}}{a}$, the cosmic scale factor $a$ is given by
\begin{equation}
a=c_{1}\left[1+\frac{3 }{4}H_{0} (3\alpha c^2 w_{h}+2 c^2 w_{h}-\alpha c^2+2)(t-t_{0})\right]^{\frac{4 }{3 (3\alpha c^2 w_{h}+2 c^2 w_{h}-\alpha c^2+2)}},
\end{equation}
where $c_{1}$ is an another positive constant of integration. One can rewrite $a$ as follows
\begin{equation}
a=a_{0}\left[1+\frac{3 }{4}H_{0} (3\alpha c^2 w_{h}+2 c^2 w_{h}-\alpha c^2+2)(t-t_{0})\right]^{\frac{4 }{3 (3\alpha c^2 w_{h}+2 c^2 w_{h}-\alpha c^2+2)}},
\end{equation}
where $a_{0}$ is the present value of the scale factor at the cosmic time $t=t_{0}$. We obtain the power-law of evolution of the Universe which avoids the big-bang singularity. \\
\indent The deceleration parameter $q$, which is defined as $q=-a\ddot {a}/\dot{a}^2$, is a geometric parameter which describes the acceleration or deceleration of the Universe depending on the negative or positive value. In this case, the deceleration parameter is given by
\begin{equation}
q=\frac{1}{2}+\frac{3}{4}c^2(3\alpha w_{h}+2w_{h}-\alpha).
\end{equation}
Here, we obtain a constant deceleration parameter as expected due to the power-law of the evolution. As we observe that for a given value of $w_{h}$, one can obtain an accelerated expansion for coupling parameter $\alpha$ satisfying the constraint $\alpha>\frac{6c^2 w_{h}+2}{3c^2(1-3w_{h})}$. For example, one can take $\alpha>-\frac{1}{3}$ for $c^2=1$ and $w_{h}=-1$ to observe the accelerated expansion. Thus, the HDE with Hubble horizon as an IR cut-off can successfully explain the accelerated expansion in the framework of $f(R,T)$ gravity without assuming the interaction between HDE and DM in contrast to the works done in [19, 55]. It is to be noted that this model does not show the phase transition as the deceleration parameter is constant.\\
\indent In order to get a robust analysis to discriminate among DE models, Sahni et al. [40, 41] have introduced a new geometrical diagnostic pair $\{r, s\}$, known as statefinder parameters, which is constructed from the scale factor and its derivatives up to the third order. The statefinder pair $\{r, s\}$ is geometrical in the nature as it is constructed from the space-time metric directly. The statefinder pair $\{r, s\}$ provides a very comprehensive description of the dynamics of the Universe and consequently the nature of the DE. It is defined as
\begin{equation}
r=\frac{\dddot a}{aH^3},\hspace{1.5cm} s=\frac{r-1}{3(q-1/2)}.
\end{equation}
In this model, we obtain the statefinder parameters $r$ and $s$ as
\begin{equation}
r=\frac{9}{8}\;c^4(3\alpha w_{h}+2w_{h}-\alpha)^2+\frac{9}{4}\;c^2(3\alpha w_{h}+2w_{h}-\alpha)+1,
\end{equation}
\begin{equation}
s=\frac{c^2}{2}\;(3\alpha w_{h}+2w_{h}-\alpha)+1.
\end{equation}
We observe that the statefinder parameters $\{r, s\}$ are constant and the values of these parameters depend on the coupling parameter $\alpha$, constant $c$ and EoS parameter $w_{h}$ of HDE. In the papers [40, 41], it has been observed that SCDM model and $\Lambda$CDM model have fixed point value of statefinder pair $\{r, s\}=\{1, 1\}$ and $\{r, s\}=\{1, 0\}$, respectively. In our work, it is observed that SCDM model can be achieved for $\alpha=\frac{2w_{h}}{1-3w_{h}}$ for any values of $c$.  Thus, for a suitable value of $w_{h}$ which may be obtained by observations, we can find the coupling parameter $\alpha$ for which $\{r, s\}=\{1, 1\}$  and viceversa. \\
\indent In a paper, Li et al. [56] have studied the Planck constraints on HDE  and obtained the tightest and self-consistent value of constant $c$ from Planck+WP+BAO+HST+lensing as $c=0.495\pm0.039$. Therefore, let us consider here and thereafter $c=0.5$ for further discussion which is lying in this observed range. Now, assuming any values of  $w_{h}$, we can get the value of coupling parameter $\alpha$, e.g., $c=0.5$ and $w_h=-1$, we have $\alpha=1.5$, which achieves the fixed point of $\Lambda$CDM model $\{r, s\}=\{1, 0\}$. Observations show that  EoS of HDE is not exactly the same of cosmological constant ($w_{h}=-1$), in fact it may lie in quintessence region ($-1<w_{h}<-1/3$) or phantom region $w_h <-1$. The coupling parameter $\alpha$ provides us the flexibility to obtain the $\Lambda$CDM  model for any values of  $w_{h}$ lying in the quintessence/phantom region. At the boundary of quintessence region i.e. $w_{h}=-1/3$ and $c=0.5$, we have $\alpha=11/3$ for which $\{r, s\}=\{1, 0\}$. Here, we observe that as we decrease the value of $w_{h}$ from -1/3 to -1, the value of $\alpha$ decreases from 11/3 to 3/2 to obtain $\{r, s\}=\{1, 0\}$. Similarly, one can get $\{r, s\}=\{1, 0\}$ in phantom region for suitable value of $\alpha$.\\

\noindent\textbf{4 Viscous holographic dark energy cosmology}\\

 \indent In non-viscous HDE model we have obtained the constant value of  deceleration parameter which does not describe the phase transition. But, the astronomical observations show that the phase transition is an integral part of the evolution of the Universe. Therefore, in this section, it would be of interest to investigate whether a viscous HDE with the Hubble horizon as an IR cut-off could be helpful to find the phase transition, i.e., time -dependent deceleration parameter and statefinder parameters in order to elucidate the observed accelerated expansion of the Universe. \\
\indent  In an accelerating Universe, it may be natural to assume that the expansion process is actually a collection of state out of thermal equilibrium in a small fraction of time due to the existence of possible dissipative mechanisms. In an isotropic and homogeneous FRW model, the dissipative process may be treated via the relativistic theory of bulk viscosity proposed by Eckart [57] and later on pursued by Landau and Lifshitz [58]. It has been found that only the bulk viscous fluid remains compatible with the assumption of large scale homogeneity and isotropy. The other processes, like shear and heat conduction, are directional mechanisms and they decay as the Universe expands. Bulk viscosity can produce an accelerated expansion even without dark energy matter due to the presence of an effective negative pressure. Recently, the present authors [37, 49] have studied the effect of viscous fluid in $f(R,T)$ gravity and discussed the recent accelerated expansion of the Universe.\\
 \indent Using the Eckart formalism for dissipative fluids [57], we can assume that the effective pressure of HDE is a sum of the thermodynamical pressure ($p_h$) and  the bulk viscous pressure ($\Pi$), i.e.,
 \begin{equation}
 P_{eff}=p_{h}+\Pi=p_h-3\zeta H,
 \end{equation}
 where $\zeta$ is the positive coefficient of the bulk viscosity. Now, the matter Lagrangian is taken as $\mathcal{L}_{m}=-P_{eff}$ for which  Eq. (4) gives $\circleddash_{\mu\nu}=-2T_{\mu\nu}-P_{eff}g_{\mu\nu}$. In this model we follow the same concept as discussed in Section 3 to analyze the behavior of the Universe. Thus, we assume that the viscous HDE matter interacts with the geometry of the Universe. Using $f(T)=\alpha T$, Eq. (3) yields the field equations for viscous HDE in the framework of $f(R, T)$ gravity as
\begin{equation}
3H^{2}=\rho_{m}+\rho_{h} + \alpha(\rho_{m}+\rho_{h}+p_{h}-3\zeta H)+\frac{1}{2}\alpha T,
\end{equation}
\begin{equation}
2\dot{H}+3H^{2}=-p_{h}+3\zeta H+\frac{1}{2}\alpha T.
\end{equation}
Using $T=\rho_h-3(p_h-3\zeta H)$ into (25) and (26), a single evolution equation of $H$ is given by
\begin{equation}
2\dot H + (\alpha+1)(\rho_{m}+\rho_{h}+p_{h})-3(\alpha+1)\zeta H=0.
\end{equation}
From (25), we get
\begin{equation}
\rho_{m}=\frac{3}{2(\alpha+1)}\;H\;[(\alpha c^2 w_{h}-3\alpha c^2-2c^2+2)H-\alpha \zeta].
\end{equation}
Now, Using (13) and (28) into (27), we get
\begin{equation}
\dot H + \frac{3}{4}(3\alpha c^2 w_{h}+2c^2 w_{h}-\alpha c^2+2)H^2-\frac{3}{4}(3\alpha+2)\zeta H=0.
\end{equation}
Eq. (29) is solvable for $H$ if the coefficient of bulk viscosity $\zeta$ is known. Many authors have studied the cosmological models by assuming the various forms of the bulk viscous coefficient (for review, see [59]). Here, we assume the bulk viscous coefficient to be a constant, i.e., $\zeta=\zeta_{0}$, which is the simplest form of $\zeta=\zeta_{0}+\zeta_{1}H$ [37, 60] by taking $\zeta_{1}=0$. Now, Eq. (29) reduces to
\begin{equation}
\dot H + \frac{3}{4}(3\;\alpha\;c^2 w_{h}+2\;c^2 w_{h}-\alpha\;c^2+2)H^2-\frac{3}{4}(3\;\alpha+2)\zeta_{0} H=0.
\end{equation}
The solution of (30) is given by
\begin{equation}
H=\frac{e^{\frac{3}{4}(3\alpha+2)\zeta_{0}t}}{c_{2}+\frac{(3\;\alpha\;c^2 w_{h}+2\;c^2 w_{h}-\alpha\;c^2+2)}{(3\alpha+2)\zeta_{0}}e^{\frac{3}{4}(3\alpha+2)\zeta_{0}t}},
\end{equation}
where $c_{2}$ is a constant of integration. The scale factor $a$ in the terms of cosmic time $t$ is
\begin{equation}
a=c_{3}\left[c_{2}+\frac{(3\alpha c^2 w_{h}+2c^2 w_{h}-\alpha c^2+2)}{(3\alpha+2)\zeta_{0}}e^{\frac{3}{4}(3\alpha+2)\zeta_{0}t}\right]^{\frac{4}{3(3\alpha c^2 w_{h}+2c^2 w_{h}-\alpha c^2+2)}},
\end{equation}
where $c_{3}>0$ is another constant of integration. Eq. (32) can be rewritten as
\begin{equation}
a=a_{0}\left[1+\frac{(3\alpha c^2 w_{h}+2c^2 w_{h}-\alpha c^2+2)}{(3\alpha+2)\zeta_{0}}H_{0}(e^{\frac{3}{4}(3\alpha+2)\zeta_{0}(t-t_{0})}-1)\right]^{\frac{4}{3(3\alpha c^2 w_{h}+2c^2 w_{h}-\alpha c^2+2)}}.
\end{equation}
 One can observe that the model avoids the big-bang singularity. In this case, the deceleration parameter is given by
\begin{equation}
q=\frac{\frac{3}{4H_{0}}[(3\alpha c^2 w_{h}+2c^2 w_{w}-\alpha c^2+2)H_{0}-(3\alpha+2)\zeta_{0}]}{e^{\frac{3}{4}(3\alpha+2)\zeta_{0}(t-t_{0})}}-1.
\end{equation}
 It is observed that the value of $q$ is  time-dependent which comes due to the introduction of bulk viscous term in HDE. The phase transition of the Universe can be explained using this value of deceleration parameter. The deceleration parameter must change it's sign from positive to negative to explain the recent phase transition (deceleration to acceleration) of the Universe. In fact, $q$ must change the sign at the time $t=t_{0}$ because we have assumed $t_{0}$ is the time where the viscous HDE begins to dominate. In other words, the Universe must decelerate for $t<t_{0}$ (matter dominated epoch) and accelerate for $t>t_{0}$ (HDE dominated epoch). We observe that the Universe shows the transition from decelerated to accelerated phase for $\alpha=-\frac{2}{3}(1+\frac{5H_{0}}{3w_{h}H_{0}-12\zeta_{0}-H_{0}})$ at cosmic time $t_{0}$. Therefore, the value of coupling parameter $\alpha$ can be obtained for a given value of $w_{h}$, which may be obtained from the observations, or vice-versa to get the transition. Thus, we have shown that the bulk viscous HDE with Hubble horizon as an IR cut-off can explain the recent phase transition of the Universe in the framework of $f(R, T)$ gravity. \\
 \indent Next, we discuss the another geometrical parameters, i.e., statefinder parameters. In this case, the statefinder parameter $r$ is obtained as
 \begin{eqnarray}
 r&=&\frac{9}{16H_{0}}[(3\alpha c^2w_{h}+2c^2w_{h}-\alpha c^2+2)H_{0}-(3\alpha+2)\zeta_{0}]\\ \nonumber
 &\times&\left[\frac{(3\alpha c^2w_{h}+2c^2w_{h}-\alpha c^2+2)H_{0}-(3\alpha+2)\zeta_{0}}{H_{0}e^{\frac{3}{2}(3\alpha+2)\zeta_{0}(t-t_{0})}}+\frac{(3\alpha c^2w_{h}+2c^2w_{h}-\alpha c^2-2)H_{0}}{e^{\frac{3}{4}(3\alpha+2)\zeta_{0}(t-t_{0})}}\right]+1.
 \end{eqnarray}
The second statefinder parameter $s$ is not given here due to complexity but one can  find it by using the values of $q$ and $r$ from (34) and (35) in $s=\frac{r-1}{3(q-1/2)}$. Our model reproduces the fixed point value $\{r, s\}=\{1, 0\}$ of $\Lambda$CDM model when the parameter $\alpha$ satisfies the condition $\alpha=-\frac{2}{3}(1+\frac{13H_{0}}{3w_{h}H_{0}-12\zeta_{0}-H_{0}})$. For this value of $\alpha$, the statefinder pair is independent of time and remains fixed throughout the evolution as in $\Lambda$CDM model. Indeed, we have obtained time-dependent statefinder pair which means that a general study  of the behaviour of this pair is needed. We plot the trajectories in $r-s$ and $r-q$ planes for some particular values of parameters $\alpha$ and $w_{h}$ to discriminate our model with existing models of DE. \\
\indent In Figs. 1a, b, the fixed points $\{r, s\}=\{1, 1\}$ and $\{r, s\}=\{1, 0\}$  have been shown  as SCDM model and $\Lambda$CDM model, respectively. It is obvious from both the figures that for any values of $\alpha$ and $w_h$, the viscous HDE model always approaches to the $\Lambda$CDM model, i.e., $\{r, s\}=\{1, 0\} $ in late the time evolution. However, in early time of the evolution our model can approach in the vicinity of SCDM model for some values of $\alpha$ as can be seen in Figs. 1a, b. It is interesting to note that there exist some negative values of $\alpha$ for which viscous HDE model starts from $\Lambda$CDM in early time and approaches to the same $\Lambda$CDM model during the late time of evolution. \\
\indent In the quiessence model with constant EoS ($Q_{1}$-model) [40, 41] and the Ricci dark energy (RDE) model [61], it has been shown that the trajectories in $r-s$ plane are vertically straight lines. In the both models, $s$ is constant throughout the evolution of the Universe, while $r$ increases in RDE model and decreases in $Q_{1}$-model starting from the initial point $r=1$. It has also been shown in [40, 41] that the trajectories for the quintessence scalar field model ($Q_{2}-$model) where the scalar potential $V(\phi)$ varies as $V(\phi)\propto \phi^{-\beta}, \beta\geq1$ approach asymptotically to the $\Lambda$CDM model in the late time. Comparing this viscous HDE model with $Q_{1}-$model and RDE model, we find that our viscous HDE model produces the curved trajectories which approach to $\Lambda$CDM model in the late time. Further, we observe that our model almost shows the similar trajectories like $Q_{2}-$model for some values of $\alpha$ and $w_h$ in $r-s$ plane, e.g., for $\alpha=-0.5$, $w_{h}=-1$ and  $\alpha=-0.4$, $w_{h}=-0.5$ as shown in Figs. 1a, b, respectively, show the similar trajectories as $Q_{2}-$model for $\beta=2$ [40, 41].\\
{\begin{center}
\begin{tabular}{cc}
\begin{minipage}{215pt}
\includegraphics[width=205pt]{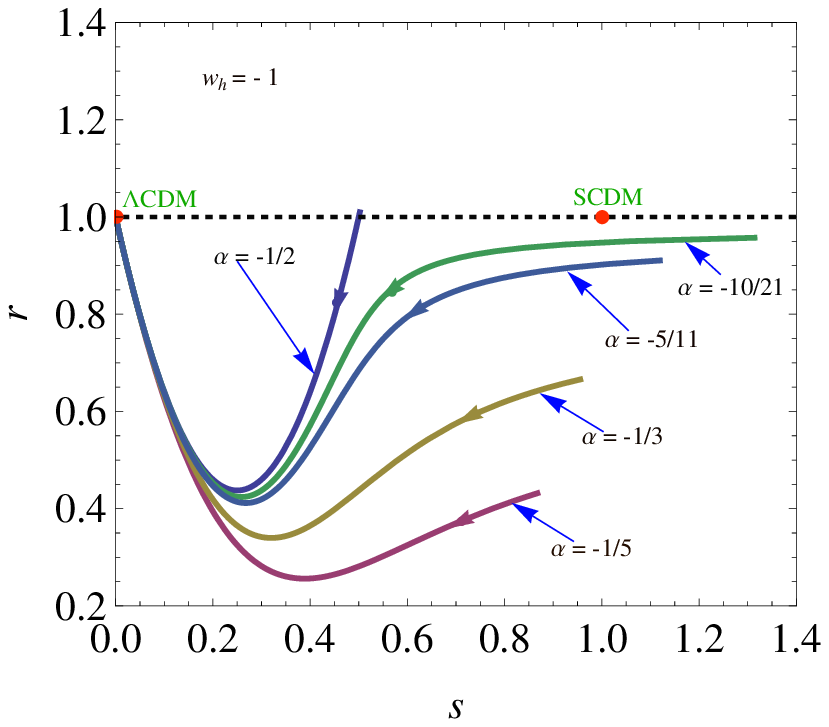}
\center{\footnotesize \textbf{(a)}}
\end{minipage}&\begin{minipage}{200pt}
\includegraphics[width=205pt]{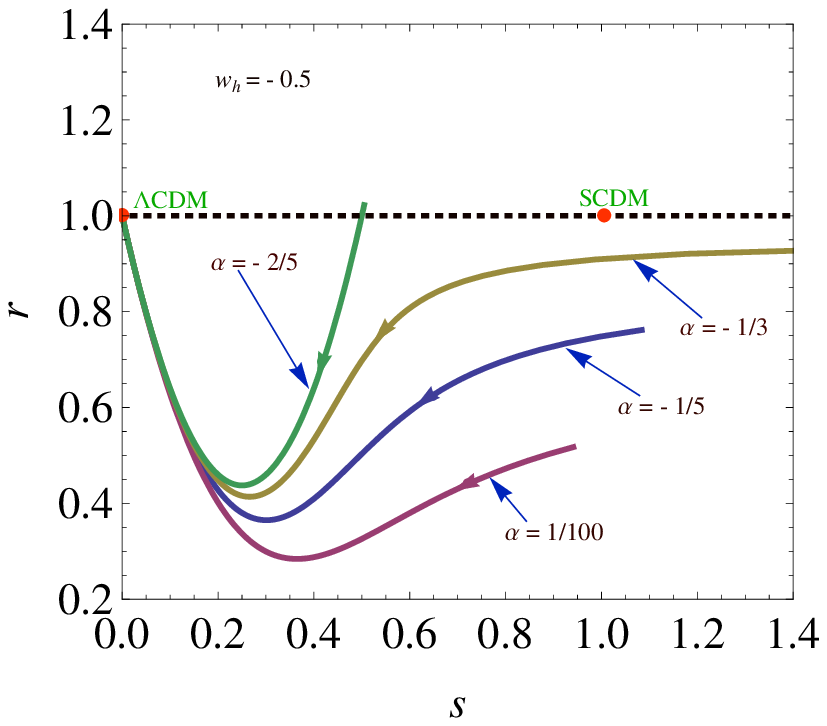}
\center{\footnotesize \textbf{(b)}}
\end{minipage}
\end{tabular}
\end{center}}
\noindent \footnotesize{\textbf{Figure 1.} The trajectories in $r-s$ plane are plotted for $w_{h}=-1$ in left panel (a) and for $w_{h}=-0.5$ in right panel (b) for different values of $\alpha$. Here, we have taken $c=0.5$, $H_{0}=1$ and $\zeta=0.05$.}\\
\normalsize

{\begin{center}
\begin{tabular}{cc}
\begin{minipage}{215pt}
\includegraphics[width=205pt]{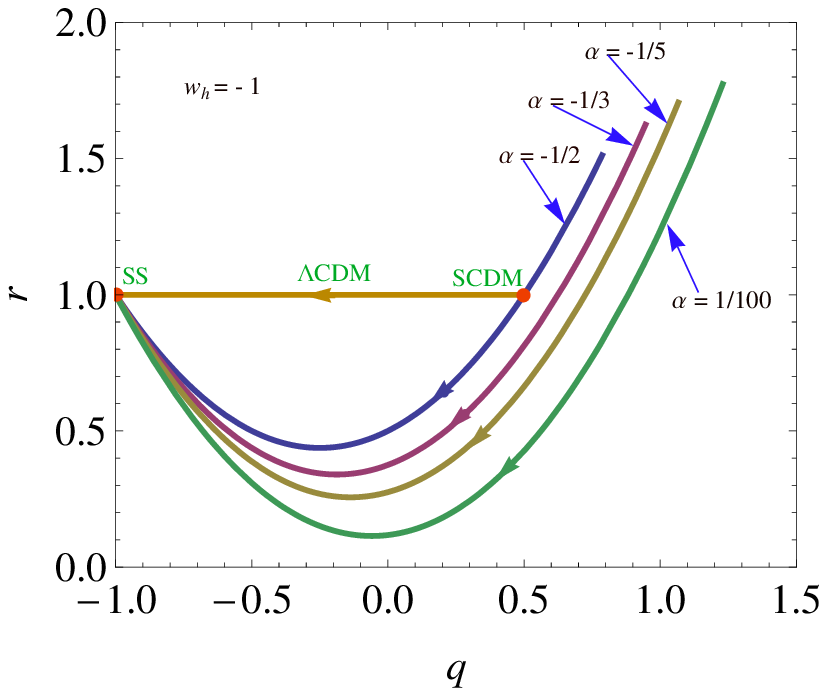}
\center{\footnotesize \textbf{(a)}}
\end{minipage}&\begin{minipage}{200pt}
\includegraphics[width=205pt]{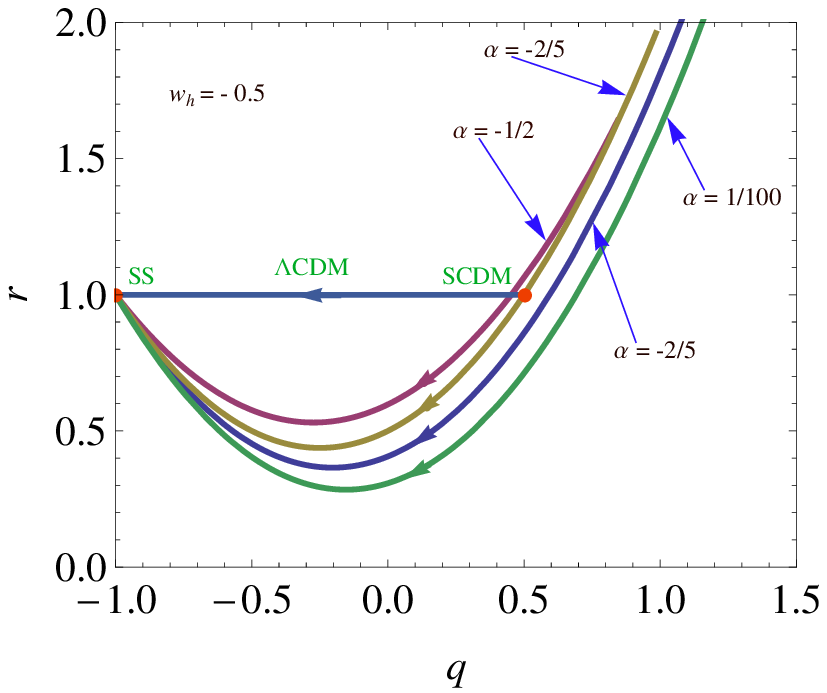}
\center{\footnotesize \textbf{(b)}}
\end{minipage}
\end{tabular}
\end{center}}
\noindent \footnotesize{\textbf{Figure 2.} The trajectories in $r-q$ are plotted for $w_{h}=-1$
in left panel (a) and for $w_{h}=-0.5$ in right panel (b) for different values
of $\alpha$. Here, we have taken $c=0.5$, $H_{0}=1$ and $\zeta=0.05$.}\\
\normalsize

 \indent Figs. 2a, b plot the trajectories in $r-q$ plane. The SCDM model and SS model (steady-state cosmology) have been shown by the fixed points $\{r, q\}=\{1, 0.5\}$ and $\{r, q\}=\{1, -1\}$ , respectively. The horizontal line in Figs. 2a, b represents the evolution of the trajectory corresponding to $\Lambda$CDM model starting from the fixed point $\{r, q\}=\{1, 0.5\}$ of SCDM model and end at the fixed point $\{r, q\}=\{1, -1\}$ of SS model. It can be observed from both the figures that for any values of $\alpha$ and $w_h$, the viscous HDE model always approaches asymptotically to the SS model, i.e., $\{r, q\}=\{1, -1\} $ as $\Lambda$CDM model in late time evolution. However, in the early time of evolution the model passes through SCDM model for some values of $\alpha$ as shown in Figs. 2a, b. The trajectories corresponding to our model start from the higher values of the pair $\{r, q\}$ but after a certain time span show the $Q_{2}$-model like behaviour. Again, comparing our viscous HDE model with $Q_{1}$-model, RDE model and $Q_{2}-$model, we find that viscous HDE model is compatible with $Q_{2}-$model. \\\\

\noindent\textbf{5 Conclusion}\\

\noindent In GR and Brans-Dicke theory, some authors [22, 55] have found that the Hubble horizon is not a viable candidate to explain the accelerated expansion of the Universe. However, Pav\'{o}n and Zimdahl [19], and  Banerjee and Pav\'{o}n [23] have shown that the interaction between HDE and DM can describe the accelerated expansion. Therefore, it is clear that one can observe accelerated expansion if the interaction between the different matter contents is considered. In this work, we have studied non-viscous and viscous HDE models with Hubble horizon as an IR cut-off in the frame work of modified $f(R, T)$ gravity. The $f(R, T)$ gravity theory presents a maximal coupling between geometry and matter. Therefore, we have explored the consequences of the coupling of matter with the geometry of the Universe instead of taking the interaction between HDE and DM as many authors have studied. However, we have assumed that only HDE of total matter (HDE+DM) couple with geometry. We have investigated the possibility whether the Hubble horizon as an IR cut-off could explain an accelerated expansion in $f(R, T)$ gravity by assuming the interaction between HDE and geometry. We have shown that the non-viscous and viscous HDE models with Hubble horizon as an IR cut-off can explain accelerated expansion in the frame work of this modified theory for suitable values of the parameters. Further, we have investigated statefinder pair $\{r, s\}$ to discriminate our non-viscous and viscous HDE models with other existing DE models. We summarize the results of these two models as follows:\\
\indent In non-viscous HDE model, we have found an accelerated expansion under the constraint of parameters. In this case, we have obtained constant deceleration and statefinder parameters. Due to constant $q$, it is not possible to analyze the phase transition of the Universe. We have found the fixed points $\{r, s\}=\{1, 1\}$ and $\{r, s\}=\{1, 0\}$ of SCDM and $\Lambda$CDM model, respectively, for suitable choice of the parameters. Thus, non-viscous HDE model is consistent with SCDM and $\Lambda$CDM models. \\
 \indent In viscous HDE model, we have obtained the recent phase transition of the Universe as the deceleration parameter comes out to be time-dependent. In this model, the statefinder parameters are the function of cosmic time $t$. These time-dependent parameters are possible due to the inclusion of bulk viscous fluid in HDE model which could explain the recent phase transition in a better way. It is interesting to note that the viscous HDE model gives the $\Lambda$CDM model fixed point $\{r, s\}=\{1, 0\}$ and remains fixed in $\Lambda$CDM model throughout the evolution for a specific value of $\alpha$ as discussed in section 3. The statefinder diagnostic have been discussed through the trajectories of $r-s$ and $r-q$ planes as shown in Figs. 1 and 2 to discriminate our model with the existing DE models. In Figs. 1a, b, it has been observed that some of the trajectories pass through the vicinity of SCDM during early time but ultimately all approach to $\Lambda$CDM model in the late time. In Figs 2a, b, it can be seen that one of the trajectories passes through SCDM model for a suitable value of $\alpha$ in early time but all the trajectories approach to SS model in the late time of evolution. It has been noticed that for some values of $\alpha$ the trajectories of the viscous HDE model are similar to the trajectories of $Q_{2}-$model [40, 41]. Therefore, the viscous HDE in the framework of $f(R, T)$ gravity gives more general results in comparison to $\Lambda$CDM and $Q_{2}-$model at least at the level of statefinder diagnostic as we are able to achieve the behavior of both the models.\\

 \noindent \textbf{Acknowledgements}\\

 \noindent We sincerely thank to Professor Tiberiu Harko for constructive comments on an earlier draft of this paper. One of the authors PK expresses his sincere thank to University Grant Commission(UGC), India for providing the grant of Senior Research Fellowship (SRF).\\

\noindent \textbf{References}\\

\noindent 1. A.G. Riess et al., Astrophys. J. {\bf659}, 98 (2007)\\
\noindent 2. N. Suzuki et al., Astrophys. J. {\bf746}, 85 (2012)\\
\noindent 3. C.L. Bennett et al., Astrophys. J. {\bf148}, 1 (2003)\\
\noindent 4. E. Komatsu et al., Astrophys. J. Suppl. {\bf192}, 18 (2011)\\
\noindent 5. W.J. Percival et al., Mon. Not. R. Astro. Soc. {\bf401}, 2148 (2010)\\
\noindent 6. P.A.R. Ade et al., Astron. Astrophys. {\bf571}, A16 (2014)\\
\noindent 7. S.M. Carroll, Living Rev. Rel. {\bf4}, 1 (2001)\\
\noindent 8. P.J.E. Peebles, B. Ratra, Rev. Mod. Phys. {\bf75}, 559 (2003)\\
\noindent 9. T. Padmanabhan, Phys. Rept. {\bf380}, 235 (2003)\\
\noindent 10. T. Chiba, T. Okaba, M. Yamaguchi, Phys. Rev. D {\bf62}, 023511 (2000)\\
\noindent 11. R.R. Caldwell, Phys. Lett. B {\bf545}, 23 (2002)\\
\noindent 12. T. Padmanabhan, Phys. Rev. D {\bf66}, 021301 (2002)\\
\noindent 13. E.J. Copeland, M. Sami, S. Tsujikawa, Int. J. Mod. Phys. D {\bf15}, 1753 (2006)\\
\noindent 14. M.C. Bento, O. Bertolami, A.A. Sen, Phys. Rev. D {\bf66}, 043507 (2002)\\
\noindent 15. M.R. Setare, Phys. Lett. B {\bf644}, 99 (2007)\\
\noindent 16. L. Xu, J. Cosmol. Astropart. Phys. {\bf09}, 016 (2009)\\
\noindent 17. A. Sheykhi, M. Jamil, Phys. Lett. B {\bf694}, 284 (2011)\\
\noindent 18. M. Li, Phys. Lett. B {\bf603}, 1 (2004)\\
\noindent 19. D. Pav\'{o}n, W. Zimdahl, Phys. Lett. B {\bf628}, 206 (2005)\\
\noindent 20. G. 't Hooft, arXiv:gr-qc/9310026\\
\noindent 21. A.G. Cohen, D.B. Kaplan, A.E. Nelson, Phys. Rev. Lett. {\bf82}, 4971 (1999)\\
\noindent 22. S.D.H. Hsu, Phys. Lett. B {\bf594}, 13 (2004)\\
\noindent 23. N. Banerjee, D. Pav\'{o}n, Phys. Lett. B {\bf647}, 477 (2007)\\
\noindent 24. K. Bamba, S. Nojiri, S.D. Odintsov, J. Cosmol. Astropart. Phys. {\bf0180}, 045 (2008)\\
\noindent 25. S. Nojiri, S.D. Odintsov, Phys. Rep. {\bf505}, 59 (2011)\\
\noindent 26. S. Nojiri, S.D. Odintsov, Phys. Lett. B {\bf631}, 1 (2005)\\
\noindent 27. A. De Felice, S. Tsujikawa, Phys. Rev. D {\bf 80}, 063516 (2009)\\
\noindent 28. K. Bamba et al., Eur. Phys. J. C {\bf67}, 295 (2010)\\
\noindent 29. E. Elizalde et al., Class. Quantum Grav. {\bf27}, 095007 (2010)\\
\noindent 30. T. Harko et al., Phys. Rev. D {\bf8}, 024020 (2011)\\
\noindent 31. M. Sharif, M. Zubair, J. Cosmol. Astropart. Phys. {\bf03}, 28 (2012)\\
\noindent 32. S. Chakraborty, Gen. Relativ. Gravit. {\bf45}, 2039 (2013)\\
\noindent 33. T. Harko, Phys. Rev. D {\bf90}, 044067 (2014)\\
\noindent 34. C.P. Singh, V. Singh, Gen. Relativ. Gravit. {\bf46}, 1696 (2014)\\
\noindent 35. E.H. Baffou, et al., Astrophys. Space Sci. {\bf356}, 173 (2015)\\
\noindent 36. H. Shabani, M. Farhoudi, Phys. Rev. D {\bf90}, 044031 (2014)\\
\noindent 37. C.P. Singh, P. Kumar, Eur. Phys. J. C {\bf74}, 3070 (2014)\\
\noindent 38. M.J.S. Houndjo, O.F. Piattella, Int. J. Mod. Phys. D {\bf21}, 1250024 (2012)\\
\noindent 39. V. Fayaj et al., Astrophys. Space Sci. {\bf353}, 301 (2014)\\
\noindent 40. V. Sahni et al., JETP Lett. {\bf77}, 201 (2003)\\
\noindent 41. U. Alam et al., Mon. Not. R. Astron. Soc. {\bf344}, 1057 (2003)\\
\noindent 42. A. Avelino, U. Nucamendi, J. Cosmol. Astropart. Phys. {\bf04}, 06 (2009)\\
\noindent 43. G.L. Murphy, Phys. Rev. D {\bf8}, 4231 (1973)\\
\noindent 44. T. Padmanabhan, S.M. Chitre, Phys. Lett. A {\bf120}, 443 (1987)\\
\noindent 45. I. Brevik, O. Gorbunova, Gen. Relativ. Gravit. {\bf37}, 2039 (2005)\\
\noindent 46. M.-G. Hu, X.-H. Meng, Phys. Lett. B {\bf635}, 186 (2006)\\
\noindent 47. C.P. Singh, S. Kumar, A. Pradhan, Class. Quantum Grav. {\bf24}, 455 (2007)\\
\noindent 48. J.R. Wilson, J.G. Mathews, G.M. Fuller, Phys. Rev. D {\bf75}, 043521 (2007)\\
\noindent 49. P. Kumar, C.P. Singh, Astrophys. Space Sci. {\bf357}, 120 (2015)\\
\noindent 50. A. Sasidharan, T.K. Mathew, Eur. Phys. J. C {\bf75}, 348 (2015)\\
\noindent 51. M. Cataldo, N. Cruz, S. Lepe, Phys. Lett. B {\bf619}, 5 (2005)\\
\noindent 52. L. Sebastiani, Eur. Phys. J. C {\bf69}, 547 (2010)\\
\noindent 53. M.R. Setare, A. Sheykhi, Int. J. Mod. Phys. D {\bf19}, 1205 (2010)\\
\noindent 54. C-J Feng, X-Z Li, Phys. Lett. B {\bf680}, 355 (2009)\\
\noindent 55. L. Xu, W. Li, J. Lu, Eur. Phys. J. C {\bf60}, 135 (2009)\\
\noindent 56. M. Li et al., J. Cosmol. Astropart. Phys. {\bf09}, 021 (2013)\\
\noindent 57. C. Eckart, Phys. Rev. {\bf58}, 919 (1940)\\
\noindent 58. L.D. Landau and E.M. Lifshitz, Fluid Machanics (Butterworth Heineman, 1987)\\
\noindent 59. R. Maartens, Class. Quantum Gravit. {\bf12}, 1455 (1995)\\
\noindent 60. A. Avelino, U. Nucamendi, J. Cosmol. Astropart. Phys. {\bf08}, 009 (2010)\\
\noindent 61. C.J. Feng, Phys. Lett. B {\bf670}, 231 (2008)\\

\end{document}